\documentclass[aps,prd,showpacs,twocolumn,floatfix]{revtex4}

\usepackage{latexsym}
\usepackage{epsfig}
\usepackage{amssymb}

\begin{document}

\title{Wormhole geometries in modified teleparralel gravity and the energy conditions}

\author{Christian G. B\"ohmer}
\email{c.boehmer@ucl.ac.uk}\affiliation {Department of Mathematics
and Institute of Origins, University College London, Gower Street,
London, WC1E 6BT, UK}

\author{Tiberiu Harko}
\email{harko@hkucc.hku.hk} \affiliation{Department of Physics and
Center for Theoretical and Computational Physics, The University
of Hong Kong, Pok Fu Lam Road, Hong Kong}

\author{Francisco S. N. Lobo}
\email{flobo@cii.fc.ul.pt} \affiliation{Centro de Astronomia e
Astrof\'{\i}sica da Universidade de Lisboa, Campo Grande, Ed. C8
1749-016 Lisboa, Portugal}

\date{\today}

\begin{abstract}
In this work, we explore the possibility that static and spherically symmetric traversable wormhole geometries are supported by modified teleparallel gravity or $f(T)$ gravity, where $T$ is the torsion scalar. Considering the field equations with an off-diagonal tetrad, a plethora of asymptotically flat exact solutions are found, that satisfy the weak and the null energy conditions at the throat and its vicinity. More specifically, considering $T\equiv 0$, we find the general conditions for a wormhole satisfying the energy conditions at the throat and present specific examples that satisfy the energy conditions throughout the spacetime.
As a consistency check, we also verify that in the teleparallel equivalent of General Relativity, i.e., $f(T)=T$, one regains the standard general relativistic field equations for wormhole physics. Furthermore, considering specific choices for the $f(T)$ form and for the redshift and shape functions, several solutions of wormhole geometries are found that satisfy the energy conditions at the throat and its neighbourhood. As in their general relativistic counterparts, these $f(T)$ wormhole geometries present far-reaching physical and cosmological implications, such as being theoretically useful as shortcuts in spacetime and for inducing closed timelike curves, possibly violating causality.
\end{abstract}


\maketitle

\section{Introduction}

Modifications of General Relativity (GR) are almost as old as the theory itself. Many of these
modifications are based on the idea of introducing additional geometrical degrees of freedom
into the theory, thereby allowing to explain observations by geometry instead of introducing
additional particles into the theory. A simple modification is the addition of a cosmological
constant which gives rise to an accelerated expansion of the universe, although several
sophisticated dark energy models have since been developed, see for instance~\cite{Durrer:2008in}.
Modified theories of gravity in which the Einstein-Hilbert Lagrangian is supplemented with additional curvature terms, have also been extensively analyzed recently~\cite{modgrav}. Although the above-mentioned modifications are based on the metric $g_{\mu\nu}$ being the dynamical variable, it is interesting to note that an alternative approach to GR has been proposed, where the tetrad $e^i{}_{\mu}$ is considered the basic physical variable. Since $g_{\mu\nu} = \eta_{ij} e^i{}_{\mu} e^j{}_{\nu}$ one is tempted to think of the tetrad as the square root of the metric, in analogy with the Dirac equation being the square root of the Klein-Gordon equation. When working with the tetrad, one naturally encounters the notion of torsion, which can be used to describe GR entirely with respect to torsion instead of curvature (without torsion) derived from the metric only. This is the so-called teleparallel equivalent of General Relativity (TEGR), see~\cite{tegr}.
This equivalent formulation can now be used as the starting point to construct modified
theories of gravity~\cite{Ferraro:2006jd,Bengochea:2008gz}. These $f(T)$ gravity models, where
$T$ is the torsion scalar, have interesting properties, for instance, the field equations are of second order, unlike $f(R)$ gravity which is of fourth order in the metric approach.

In this context, $f(T)$ models have been extensively applied to cosmology, and in particular to explain the late-time cosmic accelerating expansion without the need of dark energy~\cite{exp,Zheng:2010am,ob}. The application to cosmology is very natural in the sense that the FLRW metric is conformally flat, which makes it conceptually easier to understand in the teleparallel framework. Recently, static and spherically solutions have also been studied in the context of $f(T)$ gravity. The existence of such configurations has been establish in~\cite{Boehmer:2011gw}, and various solutions have been explicitly constructed. This construction crucially depended on an appropriate ansatz for the tetrad field, an issue which is much more delicate in the teleparallel formulations of GR than in its metric formulation. The most general static and spherically symmetric tetrads up to Lorentz transformation contain 6 free functions which results in a very complicated set of field equations which has not been studied in detail yet. Even the situation in vacuum is far from trivial as can be seen from the recent study~\cite{Ferraro:2011ks} where it turns out that a complicated tetrad is required in order to properly study the Schwarzschild solution in $f(T)$ gravity. This tetrad in general contains three free functions which become dependent in the Schwarzschild case.

In~\cite{Boehmer:2011gw}, an off-diagonal tetrad gave rise to a set of field equations which closely resembles their general relativity counterpart. A diagonal tetrad, on the other hand, resulted in highly constrained models with constant torsion scalar, which are of less physical interest in this context. The aim of the present paper is to investigate whether these $f(T)$ field equations based on the off-diagonal tetrad admit static and spherically symmetric wormhole solutions.

Wormholes are hypothetical tunnels in spacetime, through which observers may freely traverse~\cite{Morris}. In GR wormhole spacetimes are supported by ``exotic matter'',
which involves a stress-energy tensor violating the null energy condition (NEC), i.e.,
$T_{\mu\nu}k^{\mu}k^{\nu}<0$, where $k^{\mu}$ is {\it any} null vector. The issue of exotic
matter is a problematic one, so that it is useful to minimize its usage. In fact, in the
context of modified theories of gravity, it is important to note that it is the effective
stress-energy tensor that violates the null energy condition, and one may in principle allow
the normal matter threading the wormhole to satisfy the energy conditions. More specifically
in conformal Weyl gravity and $f(R)$ gravity, it is the higher order curvature terms that are
responsible for supporting the geometries~\cite{modgravity}. An observational method for the identification of wormholes through the analysis of the radiation spectrum of thin accretion disks around this type of compact objects was proposed in~\cite{thin}. In braneworlds, it was found that the local high-energy bulk effects and the nonlocal corrections from the Weyl curvature in the bulk that may induce a NEC violating signature on the brane, while the stress energy tensor confined on the brane, threading the wormhole, is imposed to satisfy the energy conditions~\cite{Lobo:2007qi}. Analogously, in $f(T)$ gravity, by considering the field equations with an off-diagonal tetrad, we find a wide range of exact solutions, where the normal matter
threading the wormhole satisfies the energy conditions, which differ radically from their general relativistic counterparts.

This paper is organized in the following manner: In Section \ref{sec1} we outline the general
formalism of modified teleparallel gravity and present the field equations with an off-diagonal tetrad. In Section \ref{sec3}, we find exact solutions of asymptotically flat wormhole geometries in $f(T)$ gravity, where the normal matter satisfy the energy conditions, in particular, the null and weak energy conditions at the throat and its vicinity. In Section \ref{concl}, we conclude.

\section{Modified teleparallel gravity}\label{sec1}

\subsection{Formalism}\label{sec1a}

In the teleparallel formulation of GR the dynamical variable is the tetrad fields $e^i{}_\mu$. The Greek indices (holonomic) denote the coordinates of the manifold while the Latin indices (anholonomic) denote the frame. We can use the same symbol to denote the inverse of $e^i{}_{\mu}$ provided we carefully stagger frame and the coordinate index. Let us
define
\begin{equation}
  e^i{}_{\mu} e_i{}^\nu = \delta^\nu_\mu \,, \qquad
  e^i{}_{\mu} e_j{}^\mu = \delta^i_j \,,
\end{equation}
where the metric is given by
\begin{equation}
  g_{\mu\nu} = \eta_{ij} e^i{}_{\mu} e^j{}_{\nu} \,.
\end{equation}
Here $\eta_{ij}=\text{diag}(1,-1,-1,-1)$ is the standard Minkowski metric, which geometrically
plays the role of the tangent space metric. The metric $g$ is used to raise and lower
coordinate indices and $\eta$ raises and lowers frame indices.

The crucial assumption of teleparallel gravity is that the manifold is globally flat. This
means that the notion of parallelism holds globally and therefore one speaks of absolute
parallelism which is a synonym of teleparallelism. This assumption is in stark contrast with GR where the notion of parallelism only holds locally and one cannot compare vectors at different points without introducing notions of transport of vectors. Due to the presence of torsion in such theories, one has to be careful when interpreting the implications of flatness. Flatness, i.e.~the vanishing of the Riemann curvature tensor (metric plus torsion), does not imply that the metric is trivial. The converse is also true, namely, one can construct geometries with non-vanishing curvature starting from a flat metric. In the latter case, it is the torsion induced by the tetrad field which will give rise to some form of curvature.

Therefore, in teleparallel gravity there exists a coordinate system where the metric is globally Minkowskian. In this case the tetrad fields give rise to a connection defined by
\begin{equation}
  \Gamma^{\sigma}_{\mu\nu} = e_i{}^{\sigma} \partial_\nu e^i{}_{\mu}
  = -e^i{}_{\mu} \partial_\nu e_i{}^{\sigma}\,,
  \label{Wbock}
\end{equation}
which is the so-called Weitzenb\"{o}ck connection. Clearly, this cannot be the Levi-Civita
connection since its torsion is zero by definition. We define torsion and contortion by
\begin{eqnarray}
  T^{\sigma}{}_{\mu\nu} &=& \Gamma^{\sigma}{}_{\nu\mu} - \Gamma^{\sigma}{}_{\mu\nu} =
  e_i{}^{\sigma} (\partial_\mu e^i{}_{\nu}-\partial_\nu e^i{}_{\mu}) \,, \\
  K^{\mu\nu}{}_{\sigma} &=& -\frac{1}{2}
  (T^{\mu\nu}{}_{\sigma}-T^{\nu\mu}{}_{\sigma}-T_{\sigma}{}^{\mu\nu})\,,
\end{eqnarray}
respectively. The contortion tensor can also be defined in terms of the Weitzenb\"{o}ck and
Levi-Civita connections. It turns out to be useful to define the tensor $S_{\sigma}{}^{\mu\nu}$ in
the following way
\begin{equation}
  S_{\sigma}{}^{\mu\nu} = \frac{1}{2}(K^{\mu\nu}{}_{\sigma} +
  \delta^\mu_\sigma T^{\rho \nu}{}_{\rho} - \delta_\sigma^\nu T^{\rho\mu}{}_{\rho})\,.
\end{equation}
Finally, we define the torsion scalar $T$ which is given by
\begin{equation}
  T = S_{\sigma}{}^{\mu\nu} T^{\sigma}{}_{\mu\nu}\,,
  \label{eqn:torsions}
\end{equation}
whose importance will become clear in a moment.

Due to the flatness of the manifold, the total Riemann curvature tensor is zero. It contains
two parts, a torsion free part defined by the Levi-Civita connection and a torsion part
expressed in terms of the Weitzenb\"{o}ck connection, or equivalently torsion. The Ricci
scalar will therefore also contain two pieces. This fact can be used to rewrite the
torsionless Ricci scalar in the Einstein-Hilbert action in terms of torsion. This particular
combination of torsion terms which appears in this context is the above mentioned $T$. Note
that the teleparallel equivalent of GR is invariant under arbitrary coordinate transformations
and local Lorentz transformations. This latter part is non-trivial to see. It suffices to note
that the teleparallel action of GR differs from its usual metric formulation only by a surface
term. As the latter is Lorentz invariant, so must be the former. The direct proof of this result is very involved and does not contain any additional information. Note that $f(T)$ gravity does
not differ from $f(R)$ by a surface term, thus local Lorentz invariance will be broken by this
theory. This also explains why the choice of tetrad is a delicate issue in $f(T)$ gravity. Two tetrads related by a local Lorentz transformation do not yield equivalent field equations related by Lorentz transformations. Different tetrads will therefore pick out specific aspects of the theory. In principle one should therefore study the field equations starting with the most general tetrad subject to the required symmetry properties of the spacetime. However, such tetrads can result is very complex field equations which cannot be solved analytically.   

Let us consider the modified action (with geometrized units $c=G=1$)
\begin{equation}
  S =
  \frac{1}{16\pi} \int e\, f(T) \, d^4x +
  \int e\, L_{\rm matter} \, d^4x \,,
  \label{eqn:action}
\end{equation}
where $e$ is the determinant of $e^i{}_\mu$.

Variations of the action~(\ref{eqn:action}) with respect to the tetrads $e^i{}_{\mu}$ gives
the field equations of $f(T)$ modified gravity which are given by
\begin{eqnarray}
  S_i{}^{\mu\nu} f_{TT} \partial_\mu T + e^{-1} \partial_{\mu}(e S_i{}^{\mu\nu}) f_T &&
     \nonumber \\
 && \hspace{-3cm}  - T^{\sigma}{}_{\mu i} S_{\sigma}{}^{\nu\mu} f_T
  - \frac{1}{4}e_i{}^{\nu}f = -4\pi \mathcal{T}_{i}{}^{\nu}\,,
  \label{eqn:field}
\end{eqnarray}
where $S_i{}^{\mu\nu}=e_i{}^{\sigma} S_\sigma{}^{\mu\nu}$, $f_T$ and $f_{TT}$ denote the first
and second derivatives of $f$ with respect to $T$, see~\cite{Bengochea:2008gz}.
$\mathcal{T}_{\mu\nu}$ is the energy momentum tensor. Conservation of the energy momentum
tensor is ensured by the field equations.

\subsection{Field equations with an off-diagonal tetrad}\label{sec2}

Consider the static spherically symmetric metric
\begin{equation}
  ds^2 = e^{a(r)} dt^2 - e^{b(r)} dr^2 -r^2 \left(d\theta^2 + \sin^2\theta\,
  d\varphi^2\right)\,,
  \label{metric1off}
\end{equation}
where $a(r)$ and $b(r)$ are two unknown functions. Following the above discussion of tetrads, we introduce the following tetrad field (one of many possible), given by
\begin{eqnarray}
  e^i{}_{\mu} = \left(
\begin{array}{cccc}
  e^{a/2} & 0 & 0 & 0 \\
  0 & e^{b/2} \sin\theta\cos\phi  & r\cos \theta\cos\phi & -r\sin \theta\sin \phi \\
  0 & e^{b/2} \sin\theta\sin\phi & r\cos \theta\sin\phi& r\sin\theta \cos\phi \\
  0 & e^{b/2} \cos \theta  & -r \sin\theta & 0
\end{array}
\right)
  \label{tetradoff}\nonumber
\end{eqnarray}
The tetrad is related to its diagonal analog by a rotation.

The determinant of $e^i{}_\mu$ is $e=e^{(a+b)/2}r^2\sin\theta$. The torsion scalar and its
derivative are
\begin{eqnarray}
  T(r) &=& \frac{2 e^{-b} \left(e^{b/2}-1\right)\left(e^{b/2}-1-r a '\right)}{r^2}\,,
  \label{defToff}\\
  T'(r) &=& -\frac{e^{-b/2}}{r^2}\left[2\left(a'-b'\right)+r \left(2a''-a'b'\right)\right]
  \nonumber \\
  &&+\frac{2e^{-b}}{r^2}\left[ \left(a'-b'\right)+r\left(a''-a'b'\right)\right]-\frac{2T}
  {r}\,,
  \label{Tscalaroff}
\end{eqnarray}
respectively. Inserting this and the components of the tensors $S$ and $T$ into the
equation~(\ref{eqn:field}) yields
\begin{eqnarray}
  4\pi\rho(r) &=& \frac{e^{-b/2}}{r}(1-e^{-b/2}) T'f_{TT}-\left(\frac{T}{4}-\frac{1}
  {2r^2}\right)f_T
  \nonumber \\
  &&+\frac{e^{-b}}{2 r^2} \left(rb '-1\right)f_T-\frac{f}{4}\,,
  \label{field:toff}\\
  4\pi p_r(r) &=&\left[- \frac{1}{2r^2}+\frac{T}{4}+\frac{e^{-b}}{2r^2}
  (1+ra')\right]f_T-\frac{f}{4}\,,
  \label{field:roff}\\
  4\pi p_t(r) &=& \frac{e^{-b}}{2}\left(\frac{a'}{2}+\frac{1}{r}-\frac{e^{b/2}}{r}\right)T'
  f_{TT}
  \nonumber \\
 &&\hspace{-2.0cm}+ f_T\left\{ \frac{T}{4}+\frac{e^{-b}}{2 r} \left[\left(\frac{1}
 {2}+\frac{ra'}{4}\right) \left(a'-b'\right)+\frac{ra''}{2}\right]\right\}-
 \frac{f}{4}\;,
  \label{field:thetaoff}
\end{eqnarray}
where $\rho(r)$ is the energy density, $p_r(r)$ is the radial pressure, and $p_t(r)$ is the pressure measured in the tangential directions, orthogonal to the radial direction.

The above field equations~(\ref{field:toff})--(\ref{field:thetaoff}) give three independent
equations for our six unknown quantities, i.e., $\rho(r)$, $p_r(r)$, $p_t(r)$, $a(r)$,
$b(r)$ and $f(T)$. This system of equations is under-determined, and we will reduce the number of unknown functions by assuming suitable conditions. Note that there is no equation enforcing the
constancy of the torsion scalar in this non diagonal gauge, contrary to the diagonal tetrad~\cite{Boehmer:2011gw}.

\section{Wormhole solutions in $f(T)$ gravity}\label{sec3}

A static and spherically symmetric wormhole is given by the metric~(\ref{metric1off}), with
the following metric function
\begin{equation}
  e ^{-b(r)} = 1-\frac{\beta(r)}{r} \,.
  \label{metricwormhole}
\end{equation}
In the context of wormhole physics $a(r)$ and $\beta(r)$ are arbitrary functions of the radial
coordinate $r$. $a(r)$ is denoted the redshift function, for it is related to the
gravitational redshift, and $\beta(r)$ is denoted the shape function, as shown by embedding
diagrams, it determines the shape of the wormhole~\cite{Morris}. The coordinate $r$ is
non-monotonic in that it decreases from $+\infty$ to a minimum value $r_0$, representing the
location of the wormhole throat, where $b(r_0)=r_0$, and then it increases from $r_0$ to
$+\infty$. To be a solution of a wormhole, one needs to impose the flaring out of the throat,
which is given by the condition $(\beta-\beta' r)/2\beta^2>0$~\cite{Morris}. At the throat we
verify that the shape function satisfies the condition $\beta '(r_0)<1$.

The flaring out condition of the throat is a fundamental ingredient in wormhole physics, and through the Einstein field equations it was found that some of these solutions possess a peculiar property, namely ``exotic matter'', involving a stress-energy tensor that violates the null energy condition. In fact, they violate all known pointwise energy conditions and averaged energy conditions. Note that the weak energy condition assumes that the local energy density is positive and states that $T_{\mu\nu}U^\mu U^\nu \geq 0$, for all timelike vectors $U^\mu$, where $T_{\mu\nu}$ is the stress energy tensor. In the local frame of the matter this amounts to $\rho>0$ and $\rho+p_i\geq0$, i.e., $\rho+p_r\geq0$ and $\rho+p_t\geq0$. By continuity, the WEC implies the null energy condition (NEC), $T_{\mu\nu}k^\mu k^\nu \geq 0$, where $k^\mu$ is a null vector.

\subsection{Specific solutions: $T(r)\equiv 0$}

As a first example, consider the specific case of $T(r)\equiv 0$. The stress-energy tensor profile is given by
\begin{eqnarray}
  4\pi \rho(r)&=&\frac{\beta'}{2r^2}f_T(0)+\frac{f(0)}{4} \,,\label{T0rho1}\\
  4\pi p_r(r)&=&-\frac{1}{2r^2}\left[1-\left(1-\frac{\beta}{r}\right)\left(1+ra'
    \right)\right]f_T(0)\nonumber\\
  &&-\frac{f(0)}{4} \,, \label{T0pr1}\\
  4\pi p_t(r)&=& \frac{1}{4r^2} \left( 1-\frac{\beta}{r} \right)\Bigg[r^2a''+
    \left(1+\frac{a'r}{2}\right)\times
    \nonumber \\
    && \left(ra'-\frac{\beta' r-\beta}{r(1-\beta/r)}\right)  \Bigg]f_T(0)-\frac{f(0)}{4}. \label{T0pt1}
\end{eqnarray}

The weak energy condition (WEC), translated by $\rho(r)\geq 0$ and $\rho(r)+p_r(r)\geq 0$ 
imposes the positivity of the right-hand-side Eq.~(\ref{T0rho1}) and of the following relationship
\begin{equation}
  4\pi [\rho(r)+p_r(r)] = \frac{1}{2r}\left[\left(1-\frac{\beta}{r}\right)a'   
  -\frac{\beta-\beta'r}{r^2} \right] f_T(0)  . \label{WEC2}
\end{equation}

The null energy condition (NEC), in addition to imposing the positivity of the left-hand-side of Eq.~(\ref{WEC2}) 
also imposes that the following condition along the tangential direction
\begin{eqnarray}
&&4\pi [\rho(r)+p_t(r)] = \frac{1}{2r^2}f_T - \frac{(\beta-\beta'r)}{4r^3}\left(1-\frac{ra'}
{2}\right)f_T  \nonumber  \\
&&  -\frac{1}{2r^2} \left(1-\frac{\beta}{r}\right)\left[1-\frac{1}{2}\left(1+\frac{ra'}{2}
\right)r a' -\frac{r^2a''}{2} \right] f_T.  \label{NEC2}
\end{eqnarray}
be positive.

It is possible to deduce specific restrictions at the throat, so that Eqs.~(\ref{T0rho1}) and~(\ref{WEC2}) evaluated at the throat reduce to
\begin{eqnarray}
  4\pi \rho|_{r_0}&=& \frac{\beta'_0}{2r_0^2}f_T(0) +\frac{f(0)}{4}\,,
 \label{pos_rho} \\
 4\pi \left(\rho+p_r \right)|_{r_0}&=&-\frac{\beta(r)-r\beta'(r)}{2r^3}\Big|_{r_0} f_T(0) \,,
\end{eqnarray}
respectively. From the flaring out condition at the throat, i.e., $(\beta-r \beta')/(2\beta^2)|_{r_0}>0$, in order to have $4\pi \left(\rho+p_r \right)|_{r_0}>0$ the condition $f_T(0)<0$ is imposed.

However, in order to have an asymptotically flat spacetime, with vanishing stress-energy components at infinity, one readily verifies from the field equations~(\ref{T0rho1})--(\ref{T0pt1}) that $f(0)=0$. This restriction implies that the
positivity of the right-hand-side of Eq.~(\ref{pos_rho}), and taking into account $f_T(0)<0$, imposes the form
function to have $\beta'_0<0$.

Further general restrictions can be deduced from Eq.~(\ref{NEC2}) evaluated at the throat,
which yields
\begin{equation}
  4\pi \left(\rho+p_t \right)|_{r_0}=\frac{1}{2r_0^2}f_T(0)-
  \frac{(1-\beta'_0)}{4r_0^2}\left(1- \frac{r_0 a'_0}{2} \right) f_T(0)\,,
  \label{NECtang}
\end{equation}
In particular, for a zero redshift function at the throat, i.e., $a_0'=0$, and imposing $4\pi
\left(\rho+p_t \right)|_{r_0} \geq 0$, we have the restriction $\beta_0'\leq -1$. For the
general case $a_0'\neq 0$, we have the condition $r_0 a_0' \leq [1-2/(1-\beta_0')]$.\\

Imposing the condition $e^{b/2}-1\neq 0$ (we refer the reader to~\cite{Boehmer:2011gw} where $e^{b/2}-1=0$ was extensively analyzed), so that Eq.~(\ref{defToff}) imposes the following first order differential equation
\begin{equation}
  \sqrt{1-\frac{\beta(r)}{r}}-1-ra'=0 \,.
  \label{odeToff}
\end{equation}

One may now consider specific choices for the shape function $\beta(r)$. For instance,
consider
\begin{equation}
  \beta(r)=\frac{r_0^2}{r} \,,
\end{equation}
so that $\beta '_0=-1$. This is the shape function considered for the Ellis wormhole~\cite{homerellis}. Thus, Eq.~(\ref{odeToff}) provides the following solution for the redshift function
\begin{equation}
  e^{a(r)}=\frac{1}{2}\left(1+\sqrt{1-\frac{r_0^2}{r^2}}  \right)\,,
\end{equation}
so that $e^{a(r)}\rightarrow 1$ as $r \rightarrow \infty$.

\begin{figure}[!ht]
\centering
  \includegraphics[width=2.6in]{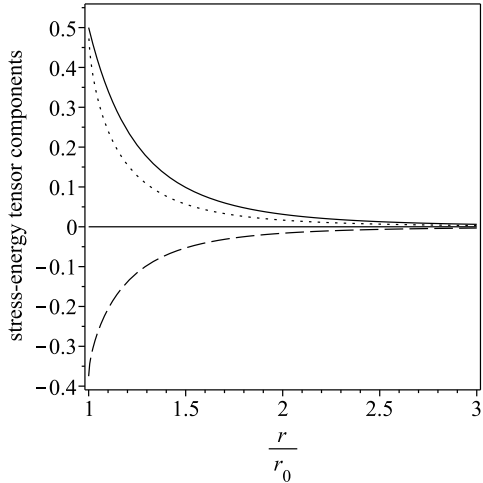}
  \caption{The figure represents the qualitative behaviour of the stress-energy components. We 
  have defined the following dimensionless quantities: $4\pi \rho(r)r_0^2/ |f_T(0)|$, $4\pi 
  p_r(r) r_0^2/ |f_T(0)|$ and $4\pi p_t(r)r_0^2/ |f_T(0)|$. The solid curve depicts the energy 
  density; the dotted the radial pressure; and the dashed curve the tangential 
  pressure.}\label{fig1}
\end{figure}

The stress-energy tensor profile for this specific case is given by
\begin{eqnarray}
  4\pi \rho(r)&=&\frac {r_0^2}{2r^4}|f_T(0)| \,,\label{T0rho2}\\
  4\pi p_r(r)&=&\frac{1}{2r^2} \left[\left(\sqrt{1-\frac{r_0^2}{r^2}}-1\right) + \frac{r_0^2}{r^2}\right]|f_T(0)|
  \,, \label{T0pr2} \\
  4\pi p_t(r)&=& \frac{1}{4r^2} \left[\left(\sqrt{1-\frac{r_0^2}{r^2}}-1\right) - \frac{r_0^2}{2r^2}\right]|f_T(0)|\,.
  \label{T0prnew}
\end{eqnarray}
The qualitative behaviour of the stress-energy tensor components is depicted in 
Fig.~\ref{fig1}, where we have defined the following dimensionless quantities: $4\pi 
\rho(r)r_0^2/ |f_T(0)|$, $4\pi p_r(r) r_0^2/ |f_T(0)|$ and $4\pi p_t(r)r_0^2/ |f_T(0)|$. Note 
that the energy density is positive throughout the spacetime due to $f_T(0) < 0$.

\begin{figure}[!ht]
\centering
  \includegraphics[width=2.6in]{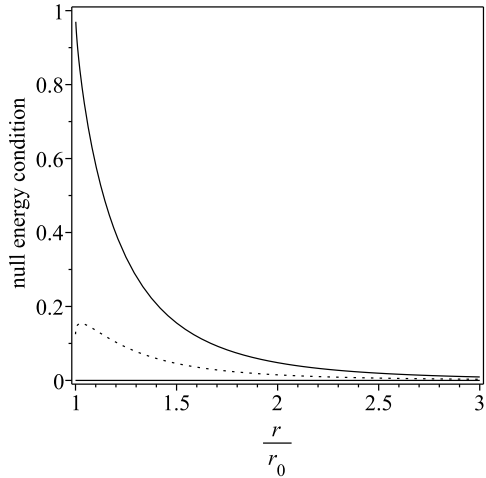}
  \caption{The figure represents the null energy condition. The 
  solid curve depicts $4\pi r_0^2(\rho+p_r)/|f_T(0)|$; the dotted curve depicts $4\pi 
  r_0^2(\rho+p_t)/|f_T(0)|$, where $f_T(0)<0$. Both quantities are positive throughout the s
  pacetime, thus satisfying the null energy condition.}\label{fig2}
\end{figure}

The NEC is given by the positivity of the left-hand-side of the following expressions
\begin{eqnarray}
  4\pi \left(\rho+p_r\right) &=& \frac{1}{2r^2} \left(1+\frac{r_0^2}{r^2} - 
  \sqrt{1-\frac{r_0^2}{r^2}}\right) |f_T(0)|, \label{T0NECr}
  \\
  4\pi \left(\rho+p_t\right) &=& \frac{1}{4r^2} \left(\frac{3r_0^2}{2r^2} - 
  1+\sqrt{1-\frac{r_0^2}{r^2}}\right)|f_T(0)|.
 \label{T0NECt}
\end{eqnarray}
The dimensionless quantities $4\pi r_0^2(\rho+p_r)/|f_T(0)|$ and $4\pi r_0^2(\rho+p_t)/|
f_T(0)|$, where $f_T(0)=-|f_T(0)|<0$, are depicted in Fig.~\ref{fig2}. In the latter it is 
transparent that the NEC is satisfied throughout the spacetime.

\subsection{Specific $f(T)$, redshift and shape functions}

\subsubsection{Teleparallel gravity: $f(T)=T$}

As mentioned in the Introduction, the tetrad can be used to describe GR entirely with respect to torsion instead of curvature. This approach is the 
so-called teleparallel equivalent of General Relativity (TEGR) \cite{tegr}. Thus, it seems 
interesting to verify if the standard general relativistic field equations of wormhole physics 
are regained from TEGR, serving as a consistency check.

Consider the specific case of teleparallel gravity, i.e., $f(T)=T$. The stress-energy tensor 
profile for this specific case is given by
\begin{eqnarray}
  4\pi \rho(r)&=&\frac{\beta'}{2r^2}\,,\\
  4\pi p_r(r)&=&\frac{1}{2r^2}\left[-\frac{\beta}{r} +\left(1-\frac{\beta}{r}\right)r a' 
  \right]\,,
  \\
  4\pi p_t(r)&=& \frac{1}{4r^2} \left( 1-\frac{\beta}{r} \right)\Bigg[\frac{r^2a''}{2}+
    \left(1+\frac{a'r}{2}\right)\times
    \nonumber   \\
    &&\left(ra'-\frac{\beta' r-\beta}{r(1-\beta/r)}\right)  \Bigg]. \label{NECtang2}
\end{eqnarray}
In order to have a positive energy density throughout the spacetime, $\beta'>0$ is
imposed. Note that these are precisely the field equations derived from general relativity 
\cite{Morris}.

The NEC along the radial direction is given by the positivity of the left-hand-side of the following relationship
\begin{equation}
  4\pi (\rho+p_r)=\frac{1}{2r^2}\left(1-\frac{\beta}{r} \right)\left[\frac{\beta-\beta' r}
    {r(1-\beta/r)}-ra'  \right] \,.  \label{NECrad}
\end{equation}
Considering that $a(r)$ is finite throughout spacetime, then one immediately finds that 
$(\rho+p_r)|_{r_0}< 0$ at the throat or at its neighbourhood, due to the flaring-out condition 
of the throat, i.e., $(\beta-\beta' r)/r^2|_{r_0}>0$. 

For the case of the null energy condition along the tangential direction, we have the
following general condition
\begin{eqnarray}
  4\pi (\rho+p_t)&=&\frac{1}{2r^2}\Bigg\{1-\frac{\beta-\beta'r}{2r}\left(1-\frac{ra'}{2}  \right)
  \nonumber  \\
  &&\hspace{-1.5cm} -\left( 1-\frac{\beta}{r} \right)\left[1-\frac{a'r}
    {2}\left(1+\frac{ra'}{2}\right)-\frac{r^2a''}{2}  \right]\Bigg\} . \label{NECtang22}
\end{eqnarray}
At the throat, this reduces to
\begin{eqnarray}
  4\pi (\rho+p_t)|_{r_0} = \frac{1}{2r_0^2}\left[1 - \frac{1-\beta_0'}{2}\left(1-\frac{r_0 a_0'}{2}
  \right) \right]\,.
\end{eqnarray}
For the general case $a_0'\neq 0$, we have the condition $r_0 a_0' \geq [1-2/(1-\beta_0')]$,
analogously to Eq. (\ref{NECtang}). These are precisely the conditions that are obtained in standard general relativity. One can now find specific solutions that have been extensively analysed in the literature, and we refer the reader to \cite{Morris}.

\subsubsection{Specific solutions: $a(r)=0$, $b(r)=r_0^2/r$ and $f(T)=T+T_0T^2$}

Consider the specific redshift function and shape function given by
\begin{equation}
  a(r)=0, \qquad b(r)=\frac{r_0^2}{r}\,,
\end{equation}
respectively, and the function $f(T)=T+T_0T^2$.

\begin{figure}[!ht]
\centering
  \includegraphics[width=2.6in]{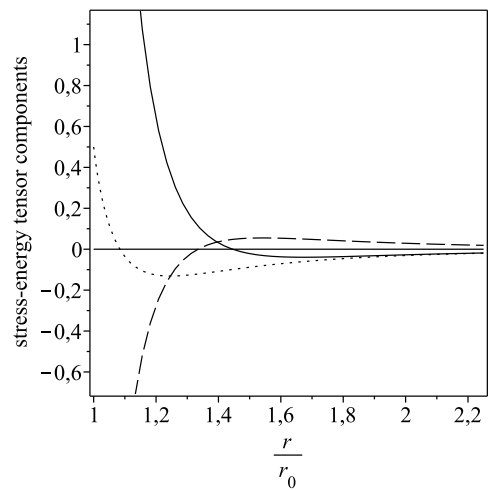}
  \caption{The solid curve depicts the energy density; the dotted the radial pressure; and the
  dashed curve the tangential pressure. We have defined the following dimensionless 
  quantities: $4\pi \rho(r)r_0^2$, $4\pi p_r(r)r_0^2$ and $4\pi p_t(r)r_0^2$. See the text for 
  more details.}\label{fig3}
\end{figure}

Inserting these functions into the stress-energy tensor profile, 
Eqs.~(\ref{field:toff})--(\ref{field:thetaoff}), provides the following expressions
\begin{eqnarray}
  4\pi \rho(r)&=&-\frac{r_0^2}{2r^4}\Bigg[1+16\frac{T_0}{r_0^2}\sqrt{1-\frac{r_0^2}
      {r^2}}\left(3-5\frac{r_0^2}{r^2} \right)
    \nonumber  \\
    &&-2\frac{T_0}{r_0^2}\left(24-52\frac{r_0^2}{r^2}+17\frac{r_0^4}{r^4}\right) \Bigg]\,,\\
  4\pi p_r(r)&=&-\frac{r_0^2}{2r^4}\Bigg[1+16\frac{T_0}{r_0^2}\sqrt{1-\frac{r_0^2}
      {r^2}}\left(1-\frac{r_0^2}{r^2} \right)
    \nonumber  \\
    &&-2\frac{T_0}{r_0^2}\left(8-12\frac{r_0^2}{r^2}+3\frac{r_0^4}{r^4}\right) \Bigg]\,,\\
  4\pi p_t(r)&=&\frac{r_0^2}{2r^4}\Bigg[1+8\frac{T_0}{r_0^2}\sqrt{1-\frac{r_0^2}
      {r^2}}\left(2-5\frac{r_0^2}{r^2} \right)
    \nonumber  \\
    &&-2\frac{T_0}{r_0^2}\left(8+24\frac{r_0^2}{r^2}-9\frac{r_0^4}{r^4}\right) \Bigg]\,.
\end{eqnarray}
The qualitative behaviour of the stress-energy components are depicted in Fig.~\ref{fig3}, 
with $T_0/r_0^2=-1$ for simplicity. We have defined the following dimensionless quantities: 
$4\pi \rho(r)r_0^2$, $4\pi p_r(r)r_0^2$ and $4\pi p_t(r)r_0^2$. Note that the energy density 
is positive at the throat, although it changes sign for a specific value of the radial 
coordinate.

\begin{figure}[!ht]
\centering
  \includegraphics[width=2.6in]{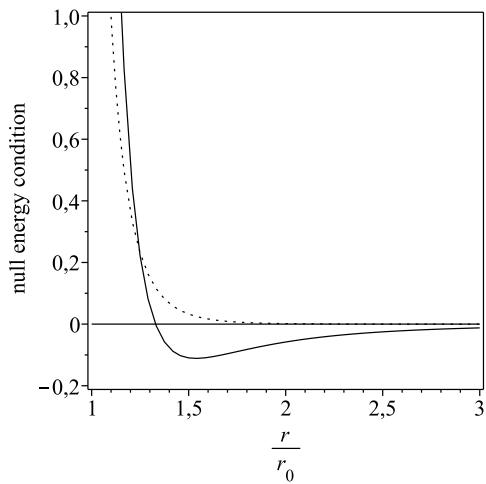}
  \caption{The solid curve depicts $4\pi r_0^2(\rho+p_r)$; the dotted curve depicts $4\pi
  r_0^2(\rho+p_t)$, and we considered that $T_0/r_0^2=-1$ for simplicity. Both quantities are
  positive at the throat and its vicinity.}\label{fig4}
\end{figure}

The NEC along the radial and the tangential directions are given by the positivity of the left-hand-side of the following expressions
\begin{eqnarray}
  4\pi [\rho(r)+p_r(r)]&=&-\frac{r_0^2}{r^4}\Bigg[1+16\frac{T_0}{r_0^2}\sqrt{1-\frac{r_0^2}
      {r^2}}\left(2-3\frac{r_0^2}{r^2} \right)
    \nonumber  \\
    &&-4\frac{T_0}{r_0^2}\left(8-16\frac{r_0^2}{r^2}+5\frac{r_0^4}{r^4}\right) \Bigg]\,,\\
  4\pi[\rho(r)+p_t(r)]&=&-\frac{4T_0}{r^4}\Bigg[\sqrt{1-\frac{r_0^2}
  {r^2}}\left(4-5\frac{r_0^2}
    {r^2} \right)
    \nonumber  \\
    &&-4+7\frac{r_0^2}{r^2}-2\frac{r_0^4}{r^4} \Bigg]\,,
\end{eqnarray}
respectively. The dimensionless quantities $4\pi r_0^2(\rho+p_r)$ and $4\pi r_0^2(\rho+p_t)$ with $T_0/r_0^2=-1$ for simplicity, are depicted in Fig.~\ref{fig4}. Note that the WEC and NEC are both satisfied at the throat and its neighbourhood, contrary to their general relativist counterparts.

\section{Summary and Discussion}
\label{concl}

General Relativity has been an extremely successful theory, with a well established experimental footing. The standard philosophy in obtaining solutions in GR is to first consider plausible distributions of matter, and through the Einstein field equation the spacetime metric of the geometry is determined. However, one may also solve the Einstein field equation in the reverse direction, namely, one first engineers an interesting spacetime metric, then finds the matter source responsible for the respective geometry. In this manner, it was found that some of these solutions possess a peculiar property, namely ``exotic matter'', involving a stress-energy tensor that violates the null energy condition. Wormhole physics is a specific example of adopting the reverse philosophy of solving the Einstein field equation. In this manner, in GR, it was found that these wormholes spacetimes possess a stress-energy tensor that violates the null energy condition. Although classical forms of matter are believed to obey these energy conditions, it is a well known fact that they are violated by certain quantum fields, amongst which we may refer to the Casimir effect and Hawking evaporation.

Thus, due to the problematic issue of exotic matter, it is useful to minimize its usage. In this context, as mentioned in the Introduction, in the context of wormhole geometries in modified theories of gravity, it is important to note that it is the effective stress-energy tensor that violates the null energy condition, and one may in principle allow the normal matter threading the wormhole to satisfy the energy conditions~\cite{modgravity,Lobo:2007qi}. In this work, we have explored the possibility that static and spherically symmetric traversable wormhole geometries were supported by modified teleparallel gravity or $f(T)$ gravity. Considering the field equations with an off-diagonal tetrad, a plethora of asymptotically flat exact solutions were found. More specifically, considering $T=0$, we found the general conditions for a wormhole satisfying the energy conditions at the throat and presented specific examples. Secondly, considering specific choices for the $f(T)$ form and for the redshift and shape functions, several solutions of wormhole geometries were found that satisfy the null and the weak energy conditions at the throat and its neighbourhood, contrary to their standard general relativistic counterparts.

However, it is rather important to further clarify some issues that were mentioned in section \ref{sec1a}. There is a very common misconception in the teleparallel equivalent of General Relativity (TEGR) which is largely based on the well known facts of (non) Riemannian geometry which no longer hold in TEGR. Namely, in GR it is well known that the vanishing of the Riemann curvature tensor implies the Minkowski metric. Thus, it is commonly said that flatness implies Minkowski. Also the converse statement is well known, namely a trivial metric with only constant coefficients implies the vanishing of the Riemann curvature tensor. Thus, when considering TEGR where the metric may be flat, one may be tempted to think that as any wormhole metric is not Riemann-flat, hence no wormholes can appear in TEGR. What then is the meaning of the solutions found in this work?

However, in teleparallel gravity the above reasoning is {\it not} true. The vanishing of the complete Riemann curvature tensor {\it does not} force the metric to be Minkowski. And the converse also holds: a trivial metric {\it does not} imply the vanishing of the Riemann curvature tensor. In particular this second statement often surprises readers and thus we provide a very simple example~\cite{Lucy}. Consider the tetrad
\begin{equation}
e^i{}_{\mu}=\left(
\begin{array}
[c]{c}%
\cos(\sqrt{x^2+y^2}) \qquad \sin(\sqrt{x^2+y^2})  \\
    -\sin(\sqrt{x^2+y^2}) \qquad \cos(\sqrt{x^2+y^2})
\end{array}
\right)   . \label{masses}%
\end{equation}

Working in 2-dimensional `Euclidean' space where the tangent space metric is simply $\delta_{ij} = \text{diag}(1,1)$, then we easily verify that the metric is of the same form $g_{\mu\nu} = \text{diag}(1,1)$, or $ds^2 = dx^2 + dy^2$, due to the trigonometric identity. However, contortion and torsion will not vanish since terms like $e_i{}^{\sigma}\partial_{\nu}e^i{}_{\mu}$ cannot vanish, thanks to the derivatives. For instance, one can confirm that the Ricci tensor component $R_{11}$ is given by
\begin{eqnarray}
  R_{11} = \frac{y (x + y \sqrt{x^2 + y^2})}{(x^2 + y^2)^{3/2}},
\end{eqnarray}
showing explicitly that this space is not flat despite having a Euclidean metric. 

There are many non-trivial spaces where the full Riemann curvature tensor vanishes and where the metric is not flat. Such spaces were first discussed mathematically by Weitzenb\"{o}ck, and are thus denoted Weitzenb\"{o}ck spaces. For as long as torsion is chosen in the correct way, the complete Riemann curvature tensor will always vanish irrespectively of the metric. 
This also implies that any topology of a given metric can always be studied in a Weitzenb\"{o}ck space instead of a Riemannian space without any torsion. As such, topological considerations are as meaningful in teleparallel gravity as they are in GR.

\section*{Acknowledgments}

The work of TH is supported by an RGC grant of the government of the Hong Kong SAR. FSNL acknowledges financial support of the Funda\c{c}\~{a}o para a Ci\^{e}ncia e Tecnologia
through the grants PTDC/FIS/102742/2008 and CERN/FP/116398/2010.

\end{document}